# Phase Based Manipulation of Airy Noise in Digital Imaging

Gajendra Singh Solanki
Mukt Mind Lab, WVC, UT, USA

*Abstract*

*Any practical imaging system, be it reflection, refraction or diffraction based, is basically band limited and therefore is bound to be affected by airy pattern noise. Apodization of the pass band is most often applied as a preferred method to get rid of these unwanted artifacts. But it comes at a cost of losing resolution as higher frequencies are tapered in apodization. Here, we propose and demonstrate a novel technique for discriminating the airy lobes based on their phase. We could thus eliminate the dominant noisy lobes without compromising with the resolution of the imaging system.*

## 1. Introduction

If phase distribution of noise and signal are different and we know how phases of signal and noises are related then it is possible to devise a phase based noise filtering (PBNF) scheme to discriminate the noise. A large number of band limited systems are affected by ringing noise or the airy pattern noise, as it is called in optical imaging. Usually, the design of a suitable windowing function is seen as its solution. Can the airy noise field in an imaging process be discriminated in phase domain and eliminated? This paper answers this affirmatively, at least for digital imaging, where it is possible to sense the phase of the imaging field, unlike real time field where frequency of light is too high to allow any device to directly sense its phase. However, use of digital imaging is rising due to ever growing computational capabilities day by day and so practical implications and applications of PBNF are abundant. Direct imaging, where some reflective or refractive optics is employed to render the image in a single step, is being replaced by indirect imaging where first a coded-image is formed and then it is digitally processed to decode the image. Digital zone plate coded incoherent holography (ZPCIH) [1-3] used here to demonstrate PBNF belongs to the latter category.

## 2. Phases of signal and noise in imaging

In the light of wide applicability of PBNF, let us treat the theory of imaging as generalized as possible. We start by representing any imaging system as a black box or a black cylinder as shown in fig 1. The finite transverse dimension of black cylinder represents the finite aperture of the

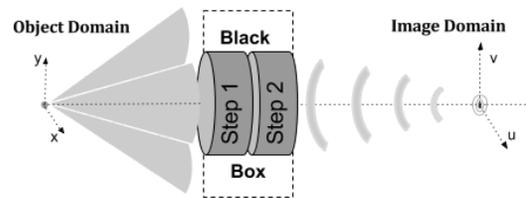

Fig 1: A generalized Model of Imaging system for direct indirect single or multi step imaging !!!

imaging system and thereby the cause of limited angular spectrum utility. The slicing of the black cylinder is optional and may represent different steps employed in imaging. Single or multi-step, the primary function of an imaging system is to reproduce every point of the object faithfully in the image domain. This demands linearity i.e. location and field strength of any point reproduced in the image should be linearly related to the corresponding point of the object. Thus, it is prudent to treat imaging as a band limited linear system. Band limited because, to produce an ideal point-image for a point-object, the imaging system must grab the whole spectrum of spatial frequencies of the point-source, but no practical system can have infinite

aperture. For a band limited linear system, the field strength in the image domain can be readily written as:

$$I(u,v) = \int\int_{-\infty}^{\infty} h(u-Mx, v-My) \cdot [\tfrac{1}{M} O(x,y)] \, dx\,dy \qquad (1)$$

where $O(x,y)$ is the field distribution in object domain, $M$ is the magnification factor of the imaging system, $h(u,v)$ is the response of imaging system to a point-object and is called impulse response (IR) or point spread function (PSF). For a diffraction limited system, PSF can be safely taken as diffraction pattern of its entrance pupil and thus,

$$h(u,v) = \frac{A.\exp(jkz)}{j\lambda z} \cdot \exp[jk.\tfrac{(u^2+v^2)}{2z}] \int\int_{-\infty}^{+\infty} P(x,y).\exp[jk.\tfrac{(ux+vy)}{z}]dx\,dy$$

$$\simeq \frac{A.\exp[jk.\tfrac{(u^2+v^2)}{2z}]}{\lambda z} \cdot FT[P(x,y)] \qquad (2)$$

where A is constant Amplitude, $P(x,y)$ is the transparency function of the entrance pupil and $FT$ stands for its Fourier transform, the very origin of lobes. The sidelobes of $h(u,v)$ give rise to noise and the central lobe acts as the imaging signal.

Our interest is to decode the phases for both the signal and the noise. Any constant phase term that affects all lobes equally is ignored. Spectral components of an $FT$ split into two groups: in-phase and out-of-phase. By combining the same with the phase curvature term outside the $FT$ integral, the desired phase distribution is obtained and clearly depicted in fig 2(a). Here each patch of solid curves gives the phase of corresponding lobe of the IR. A typical IR field is shown in fig 2(b). The effect of phase curvature is visible in (a) but not in (b). Thus, fig 2(a) forms the basis of our PBNF scheme as it not only shows the split of odd and even orders in phase domain, but also that the phases of lobes go on increasing with their order number. Summarily, we derive three different modes of operation of PBNF. **1.** If we tune PBNF module to discard contribution from a negative phase field then odd number lobes should vanish **2.** If we tune PBNF just to retain the base phase of the central lobe then all sidelobes should vanish. **3.** Even within a lobe, phase distribution is not flat but each possesses a finite continuous phase-spread. Each of these capability of PBNF shall be demonstrated in the next section. We end this section by noting the role of convolution integral of eq (1). It not only spreads the airy pattern noise of PSF throughout the image especially near sharp edges but also acts as a recipe for airy noise build up, which will also be evident from our imaging results.

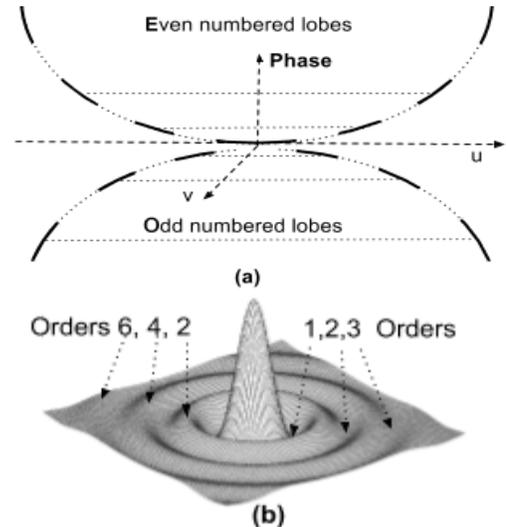

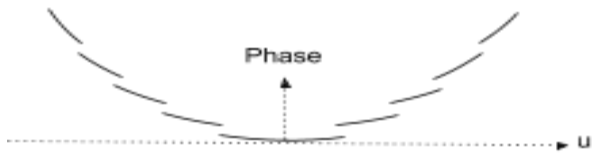

Fig 2: (a) Phase information coded in impulse response. (b) A typical field distribution for a band limited imaging system.

## Phase Based Manipulation of Airy Noise

From the findings of fig 2(a), the phase of $n^{th}$ order lobe about a point $(u_0,v_0)$ can be written as:

$$\phi_n(u,v) = \phi_0 + \pi n + \pi/z\lambda\,[(u-u_0)^2 + (v-v_0)^2] \qquad (3)$$

Where $\phi_0$ is the base phase of the image signal and depends on imaging technique, for example $\phi_0 = \pi/2$ for the first order image of ZPCIH, $\lambda$ is the wavelength of radiation. The second term distinguishes the phase of one lobe from another. The third term gives the phase variation within a lobe. Eq (3) is the basis of PBNF.

In digital ZPCIH, the grabbed digital hologram is digitally decoded by simulating coherent optical reconstruction [2,3]. So, the optical fields contributing to image and their phases are readily available to the PBNF module to make a decision to accept or reject it. Let us start by demonstrating how airy noise can severely build up and even shoot over the signal strength. Fig 4 shows the result of imaging a tiny annular ring. The first order noise at the center has added up to shoot over the signal strength. Intensity based processing would have failed to identify this fake peak as noise but PBNF is not fooled. Left image of fig 5 shows the capability of PBNF to filter out not only the first order, but also all the odd orders. Intensity plot confirms the same. Finally, in the right image, the sidelobes of all orders have been filtered out but zeroth order phase is retained. In the results of fig 5 we demonstrate the manipulation based on phase spread within a given lobe. Object H is imaged to give the top middle noisy image. The last image is cleaned of all sidelobe-noise by PBNF. But the three images in between contain the second order noise in different extent. PBNF in these images is tuning the phase within phase-spread of the second lobe to allow its different extents. This confirms the phase distribution within a lobe is not flat. Next, we consider more complex object to image. In fig 6, the object 'MUKT', its noisy image, odd order filtered image and the final PBNF cleaned image follow in order. Lastly, in fig 7, the imaging result of using a Gaussian apodizer to suppress the Airy noise is compared with the PBNF enabled imaging-result. As Gaussian apodizer suppresses the higher frequencies of the passband therefore it loses resolution and signal-strength compared to the PBNF method, which utilizes all spectral components within the passband. The intensity curves along a line passing from the middle of each image confirms both: 1. deteriorated resolution and 2. reduced signal strength.

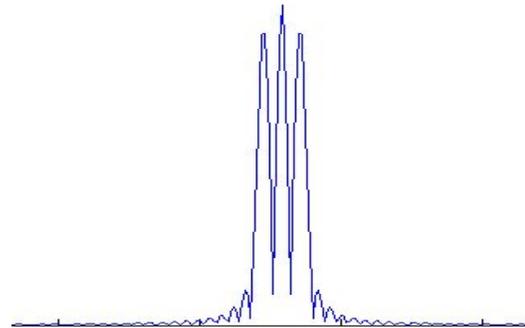

Fig 4(a) Ring-object (left), its noisy image (right). Intensity plot along a line passing through the centre of the noisy image.

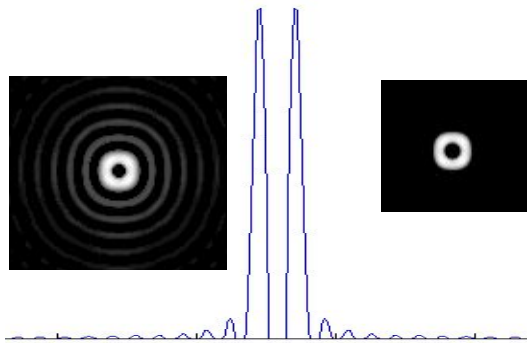

Fig 4(b) Image of ring with odd orders removed (left), its intensity plot (middle), The last image is made free from all sidelobes.

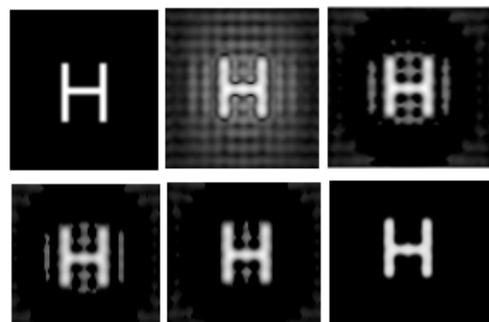

Fig 5. Top left is object then noisy image. Top-right to bottom-middle are the images with different extent second order noise. The last is free from all sidelobes.

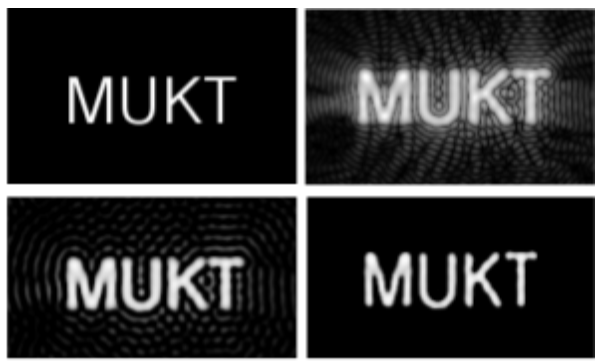

Fig 6: the object (top-left) with its noisy image on its right. PBNF applied to remove odd order noise (bottom-left) and finally both odd and even order noise is filtered out.

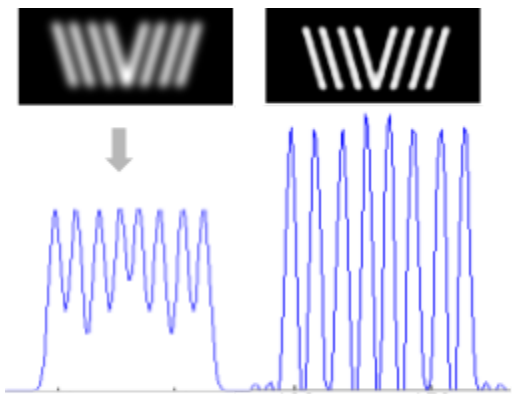

Fig 7. Left image is the result of using Gaussian apodizer. Right one is the result of PBNF for odd order noise elimination. Below each are the respective intensity plots along a horizontal line through centre.

**4. Conclusion**

Ringing noise affects a wide variety of band limited linear systems. Windowing and intensity based processing are the usual tools to tackle them. For the first time, to the best of our knowledge, a phase based solution to manipulate airy noise has been devised and successfully demonstrated for digital imaging. PBNF is superior over gaussian windowing in terms of resolution and signal strength. Applications of digital imaging in general and digital ZPCIH in particular are wide, right from holography [1], tomography [4], astronomy, plasma [5,6] and nuclear radiation [7] to highly computation driven medical imaging techniques. As a future scope this concept can be extended to advanced digital image processing, signal processing, lens imaging and CT scan tomographic projections which are often marred with ringing noise.